\documentclass[iop]{emulateapj}
\shortauthors{Sheets \& Deming}
\shorttitle{Statistical Eclipses of {\it Kepler} Sub-Saturns}
\begin{document}
\title{Statistical Eclipses of Close-in {\it Kepler} Sub-Saturns}

\author{Holly A. Sheets\altaffilmark{1,2} and Drake Deming\altaffilmark{1,2}}
\altaffiltext{1}{Department of Astronomy, University of Maryland, College Park, MD 20742-2421}
\altaffiltext{2}{NASA Astrobiology Institute's Virtual Planetary Laboratory}
\email{hsheets@astro.umd.edu}

\keywords{methods:  data analysis --- occultations --- planetary systems --- techniques:  photometric --- planets and satellites:  surfaces --- planets and satellites:  atmospheres}   

\begin{abstract}
We present a method to detect small atmospheric signals in {\it Kepler}'s planet candidate light curves by averaging light curves for multiple candidates with similar orbital and physical characteristics.  Our statistical method allows us to measure unbiased physical properties of {\it Kepler}'s planet candidates, even for candidates whose individual signal-to-noise precludes the detection of their secondary eclipse.  We detect a secondary eclipse depth of 3.83$^{+ 1.10}_{- 1.11}$ ppm for a group of  31 sub-Saturn ($R < 6 R_{\oplus}$) planet candidates with the greatest potential for a reflected light signature (($R_p/a)^2 > $ 10 ppm).  Including Kepler-10b in this group increases the depth to 5.08$^{+ 0.71}_{- 0.72}$ ppm.   For a control group with ($R_p/a)^2 < $ 1 ppm, we find a depth of 0.36 $\pm$ 0.37 ppm, consistent with no detection.  We also analyze the light curve of Kepler-10b and find an eclipse depth of 7.08 $\pm$ 1.06 ppm.  If the eclipses are due solely to reflected light, this corresponds to a geometric albedo of 0.22 $\pm$ 0.06 for our group of close-in sub-Saturns, 0.37 $\pm$ 0.05 if including Kepler-10b in the group, and 0.60 $\pm$ 0.09 for Kepler-10b alone.   Including a thermal emission model does not change the geometric albedo appreciably, assuming $A_B = (2/3)*A_g$.  Our result for Kepler-10b is consistent with previous works.  Our result for close-in sub-Saturns shows that Kepler-10b is unusually reflective, but our analysis is consistent with the results of \citet{demory} for super-Earths.  Our results also indicate that hot Neptunes are typically more reflective than hot Jupiters.
\end{abstract}

\section{Introduction}
Secondary eclipses have been detected in {\it Kepler} data for hot Jupiters \citep[e.g.][]{coughlin, esteves}, showing that these bodies have very low albedos, as predicted by atmospheric models \citep{sudarsky}.   For super-Earth-sized planets, eclipses were first detected with {\it Kepler} in the two extremely hot, close-in planets Kepler-10b \citep{kep10} and Kepler-78b \citep{sanchis}.  Unlike the hot Jupiters, these two planets show relatively high geometric albedos, between 0.3 and 0.6.  These planets are unlikely to harbor substantial atmospheres at such extreme temperatures ($>$ 1500 K) and are possibly a new class of ``lava ocean'' planets \citep{leger2011,rouan}.  The {\it Kepler} data set contains many super-Earths and sub-Saturn-sized candidates at slightly less extreme temperatures.  At slightly lower temperatures and greater distances from the host star, the eclipse signals from these candidates are much weaker.   \citet{demory} constrains the geometric albedos for 27 super-Earth candidates in the {\it Kepler} catalog and finds that the albedos for these candidates are statistically larger than those of hot Jupiters (0.16 to 0.30 versus 0.06 to 0.11), with a subset of unusually bright candidates like Kepler-10b, with albedos greater than 0.4.  Only a few of the candidates \citet{demory} analyzed show significant eclipses, while the remainder of the sample had upper limits set for their albedos.  The hierarchical Bayesian analysis of \citet{demory} also illustrates the potential for extracting information on the atmospheres of {\it Kepler}'s planets using statistical techniques.

In this paper, we introduce a statistical method wherein we average the photometric data from different planetary candidates, after linearly scaling their orbital phases so that their eclipses have the same temporal cadence.  This method has two advantages.  Our grand-average eclipse has a much higher signal-to-noise than even the stacked eclipses of individual planets, and we avoid the selection bias that could result from focussing on the most easily measured individual planets.  We focus on objects of less than 6 Earth radii.  We choose this cutoff in radius to study the smallest planets while still providing a large sample size. 

This paper is organized as follows.  Sec. 2 describes how we select Kepler data for candidate planets, and how we process those data to obtain high fidelity for our grand average eclipses.  Sec. 3 describes how we transform the individual eclipses so that they reinforce in the grand average, and Sec. 4 calculates models of reflected light and thermal emission for comparison to our results.  Sec. 5 gives our results for our candidate list of sub-Saturn sized planets (5.1), for a control group (5.2), and it briefly discusses notable individual planets (5.3), and acknowledges the limitation of our technique (5.4).  Sec. 6 discusses the implications of our results for atmospheres and surfaces of sub-Saturn sized planets, and Sec. 7 summarizes our conclusions.

\section{{\it Kepler} Observations}
We select two groups of candidates with radii less than 6 $R_{\earth}$ from the NASA Exoplanet Archive's Kepler Objects of Interest table\footnote{http://exoplanetarchive.ipac.caltech.edu/}, downloaded on December 17, 2013.  
We select the groups based on the expected depth of the candidate's eclipse, if the eclipse was due solely to reflected light with a geometric albedo of 1.  The depth of the eclipse is given by the albedo times $(R_p/a)^2$, where $R_p$ is the radius of the planet and $a$ is the orbital radius of the planet (cf. Sec. 4).  The first group consists of 32 candidates with $(R_p/a)^2 >$ 10 ppm, which makes them more likely to be detectable even with low albedo.  The objects in this group and their parameters are listed in Table \ref{tab:params}.   The second list of 376 candidates is the control list, with $(R_p/a)^2 <$ 1 ppm, which makes them undetectable. \citet{fressin} determined that 90 percent of the planet candidates are likely real planets across all radii, with smaller planets having slightly better odds of being real.  We use the short-cadence ($\approx$ 60 s exposure) PDC data from the Mikulski Archive for Space Telescopes (MAST\footnote{http://archive.stsci.edu/kepler/}), for quarters 0 through 16.  We initially normalize each light curve file for a given candidate by the mean flux outside of transit.   We check if there are multiple candidates in the system, and, if so, we set to NaN all the transits from the other candidates.  We extract each individual eclipse from the light curve file, and we include up to $\pm$ 3 durations of the eclipse.  For each extracted eclipse, we refine the normalization by fitting a line across the out-of-eclipse section to determine the baseline and dividing by the fit.  We assume the orbits are circular, and therefore the eclipses are centered at phase 0.5.   We expect these planets to have circularized, based on the finding of \citet{kane} that smaller {\it Kepler} candidates have a low mean eccentricity.  We use a 3-$\sigma$ clip to exclude unusually high or low points. We include only eclipses with at least 20 points out of eclipse on each side of the eclipse, as well as 40 points within eclipse.  We apply this cut to ensure that we are adding eclipses with a well-determined out-of-eclipse baseline level and with information during eclipse.  

The {\it Kepler} light curves occasionally contain discontinuities and ramps in flux due to systematic effects from the spacecraft and sudden changes in pixel sensitivity \citep{datacharhand}.  We use the quickMAP PDC light curves \citep{pdca, pdcb, drn21}, which have largely been corrected for these effects.  The PDC pipeline does not always catch every anomaly, however, so we must screen the data before use.  We check each individual, normalized eclipse by applying three tests, which we refer to as the projection test, the slope test, and the red noise test.

In the projection test, we fit a line to the points before ingress, and then calculate the projection of that line for the points after egress.  We take the mean of the actual data points after egress and a mean of the projected points from the linear fit.   If the means differ by more than 0.001, the eclipse is dropped.  This value was determined by trial and error, to balance the necessity of eliminating poorly fit eclipses, while maintaining a significant sample size.  We repeat the test, fitting a line to the points after egress and projecting that fit to the points before ingress.  The eclipse is dropped if it fails either part of this test;  it need not fail both parts.  This test eliminates eclipses in which there is an offset in the out of eclipse baseline due to instrumental effects.  It also helps to eliminate eclipses in cases where the approximation of a linear baseline across the eclipse was not adequate for the normalization.

In the slope test, we compare the slopes of the two lines fitted during the projection test.  We expect the slopes to be zero, since we have already normalized with a straight line.  If both slopes are consistent with zero within the 3-$\sigma$ uncertainties on the fits, the eclipse is kept.  Since the noise in any single eclipse is quite large, this test is not sensitive enough to accidentally eliminate eclipses due to phase variations.   

We implement the red noise test by calculating the standard deviation of the light curve section under consideration when it is binned by 3, 5, 7, 9, 11, 13, and 15 points.  Random noise, like photon-counting, should follow a log($\sigma$) $\alpha$ -0.5$*$log(N) relation.  For each individual eclipse, we fit a line to log($\sigma$) versus log(N).  We expect the distribution of the slope of this fit to be a Gaussian distribution around -0.5, if the noise in the eclipses is random noise.  We compile the distribution of fitted slopes for all eclipses in the group being averaged that pass the projection test.  We then fit a Gaussian distribution to the histogram of slopes and estimate the cutoff value at which the distribution appears to deviate from the Gaussian fit.  We then eliminate any individual eclipse with a slope greater than the cutoff.  For our group of close-in candidates, this cutoff slope value is -0.30, while for the control group, it is -0.35.  We eliminated KOI-3.01 (a.k.a Kepler-3b, HAT-P-11b) and KOI-2276.01 from the close-in group entirely because these two candidates have a very high rate of failure for the red noise test.  For KOI-3.01, 177 of 207 eclipses (85.5 \%) fail this test, while for KOI-2276.01, 234 of 319 eclipses (73.4 \%) fail.

\section{Averaging Candidate Light Curves}
\begin{figure}
\epsscale{0.95}
\vspace{0.1 in}
\hspace{0.3 in}
\plotone{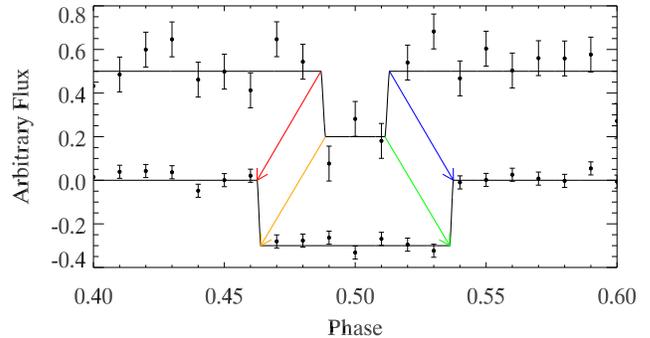}
\vspace{0.4 in}
\caption{A visualization of the rescaling of the phase to the reference object, using simulated data.  The top light curve is a single eclipse from one candidate, which is rescaled to the phase of the reference object and added into the group average, shown in the bottom light curve.  The phase of the single eclipse is broken into segments:  0.25 to 1st contact, 1st contact to 2nd contact (ingress), 2nd contact to 3rd contact (full eclipse), 3rd contact to last contact (egress), and last contact to 0.75. }
\label{fig:scaling}
\end{figure}
To constructively add the eclipses of multiple objects, we adopt the candidate from the group being averaged with the largest duration eclipse in phase to serve as a reference object.  Then we transform the phase of all other objects in the group by scaling sections of the light curve phase for each candidate to the reference object.  Figure \ref{fig:scaling} shows how we break up the light curve into segments for scaling.   Fixed points for the transforms are phase equal to 0.25, 0.5, and 0.75,  i.e. a phase of exactly 0.25, 0.5, or 0.75 is transformed to be the same in the original coordinate and the transformed coordinate.  The other reference points are 1st, 2nd, 3rd, and 4th contacts.  We choose these points so that ingress, egress, and full eclipse will add constructively in the transformed phase.  We also impose the cutoffs at 0.25 and 0.75 in phase because, for a simple circular orbit with reflected light, these points are where the inflection point for the phase curve of the planet would be located.  We transform the phase using the following equations:
\begin{equation} 
\psi = \psi_{1st} + (\phi-\phi_{1st})*\frac{\psi_{1st}-0.25}{\phi_{1st}-0.25},~~~~~~~~(0.25 < \phi < \phi_{1st})
\end{equation}
\begin{equation} 
\psi = \psi_{2nd} + (\phi-\phi_{2nd})*\frac{\psi_{2nd}-\psi_{1st}}{\phi_{2nd}-\phi_{1st}},~~~~~~~~(\phi_{1st} < \phi < \phi_{2nd})
\end{equation}
\begin{equation} 
\psi = 0.5 + (\phi-0.5)*\frac{0.5-\psi_{2nd}}{0.5-\phi_{2nd}},~~~~~~~~(\phi_{2nd} < \phi < 0.5)
\end{equation}
\begin{equation} 
\psi = 0.5 + (\phi-0.5)*\frac{\psi_{3rd}-0.5}{\phi_{3rd}-0.5},~~~~~~~~(0.5 < \phi < \phi_{3rd})
\end{equation}
\begin{equation} 
\psi =  \psi_{3rd}+ (\phi-\phi_{3rd})*\frac{\psi_{4th}-\psi_{3rd}}{\phi_{4th}-\phi_{3rd}},~~~~~~~~(\phi_{3rd} < \phi < \phi_{4th})
\end{equation}
\begin{equation} 
\psi = \psi_{4th} + (\phi-\phi_{4th})*\frac{0.75-\psi_{4th}}{0.75-\phi_{4th}},~~~~~~~~(\phi_{4th} < \phi < 0.75)
\end{equation}
where $\psi$ is the transformed phase, $\phi$ is the native phase for the object, $\psi_{1st},\psi_{2nd},\psi_{3rd}$, and $\psi_{4th}$ are the contact points in phase for the reference object, and $\phi_{1st},\phi_{2nd},\phi_{3rd}$, and $\phi_{4th}$ are the contact points in the native phase of the object.  The contact points, assuming circular orbits, are calculated using Equations 14 and 15 from \citet{winn}:
\begin{equation}
T_{tot} = \frac{P}{\pi}\mathrm{sin}^{-1}\left[\frac{R_*}{a}\frac{\sqrt{(1+k)^2-b^2}}{\mathrm{sin}i}\right]
\end{equation}
\begin{equation}
T_{full} = \frac{P}{\pi}\mathrm{sin}^{-1}\left[\frac{R_*}{a}\frac{\sqrt{(1-k)^2-b^2}}{\mathrm{sin}i}\right]
\end{equation}
where $T_{tot}$ is the time between first and last contact, $T_{full}$ is the time between second and third contact, $b$ is the impact parameter, $P$ is the period of the planet, $i$ is the inclination of the planet's orbit, $a$ is the orbital distance of the planet, $R_*$ is the radius of the star, and $k = R_p/R_*$ with $R_p$ being the radius of the planet.   We use $b$, $P$, $a$, $R_*$, and $R_p$ from the Exoplanet Archive candidate table and calculate $i$ and $R_p/R_*$ from those values.
We then bin the normalized flux data using this scaled phase coordinate and average the points in each bin, weighted by their photometric errors.  The bin size is chosen such that there are 11 bins within full eclipse.   Ingress and egress are each given their own bin, as well, which results in a different bin size for these two bins.

\section{Reflected and Thermal Light Modeling}
We calculate the reflected light contribution, given by:
\begin{equation}
\frac{F_p}{F_*}=A_g\left(\frac{R_p}{a}\right)^2
\end{equation}
where $F_p$ is the reflected light of the planet, $F_*$ is the flux from the star, $A_g$ is the geometric albedo, $R_p$ is the planet radius, and $a$ is the orbital distance of the planet.  

To calculate the thermal emission, we first estimate the effective temperature of the planet, as in, e.g. \citet{esteves}, \citet{lopez}, and \citet{rowe06}:
\begin{equation}
T_p = T_*\left(\frac{R_*}{a}\right)^{1/2}[f(1-A_B)]^{1/4}
\end{equation}
where $T_p$ is the temperature of the planet, $T_*$ is the effective temperature of the star, $a$ is the orbital distance of the planet, $R_*$ is the radius of the star, $A_B$ is the Bond albedo.  The reradiation factor $f$ ranges from 1/4 for complete redistribution of heat around the planet to 2/3 for instant reradiation.  We calculate the thermal emission of the planet assuming a blackbody spectrum, integrating over the wavelength range of {\it Kepler}, accounting for its transmission function\footnote{http://keplergo.arc.nasa.gov/kepler\_response\_hires1.txt}.  We adopt $A_B$ = $(3/2)A_g$ for a Lambert sphere, as in \citet{esteves} and \citet{lopez}, and we do this calculation for the cases of $f = 1/4$ and $f = 2/3$, with $A_g = $ 0.0, 0.1, 0.3, and 0.6.  We then normalize by the stellar flux and multiply by $(R_p/R_*)^2$.  We calculate the stellar flux for the close-in group by integrating over an ATLAS\footnote{http://kurucz.harvard.edu/} model atmosphere \citep{kurucz} for the stellar effective temperature of each candidate's host, modified by the instrument's transmission function.  For the control group, we integrate over a blackbody spectrum with the stellar effective temperature, rather than a model atmosphere.  We use the planetary and stellar parameters from the NASA Exoplanet Archive candidate table, which includes the revised effective stellar temperatures from \citet{pinson} and \citet{spc}.  Some of the KOIs in the candidate table also have updated effective temperatures based on the methods of  \citet{sme}.

\section{Results and Discussion}
\subsection{Close-in Candidates}
\begin{figure}
\epsscale{0.95}
\vspace{0.2 in}
\hspace{0.5 in}
\plotone{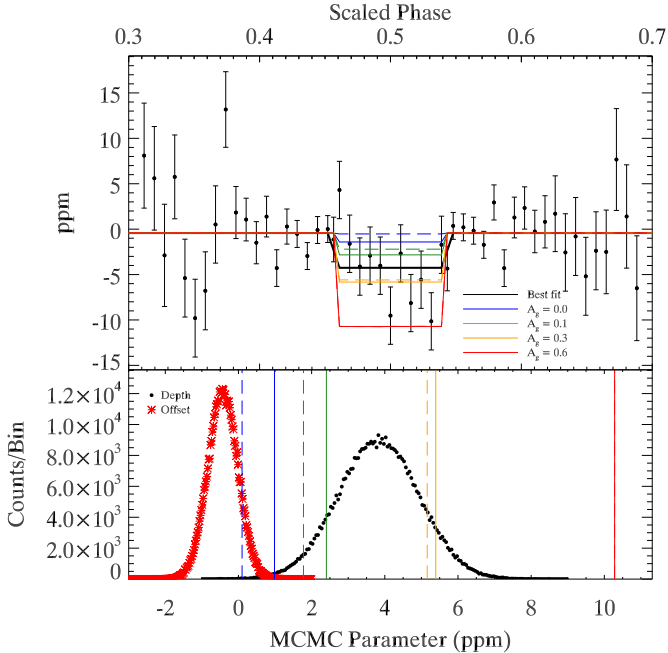}
\vspace{0.4 in}
\caption{Upper panel:  The light curve, centered on secondary eclipse, for the group of objects less than 6 R$_{\earth}$ with $(R_p/a)^2 >$ 10 ppm, excluding Kepler-10b.  The binned data are shown as points.  The error bars are the propagated photometric errors.  The best fit curve is the solid black line.  Overplotted are the reflected light plus thermal emission models for $A_g = (2/3)*A_B =$ 0.0 (blue), 0.1 (green), 0.3 (orange), and 0.6 (red), with the re-radiation factor $f =$ 1/4 (dashed) and 2/3 (solid).   Lower panel:  The distributions for the two parameters of the MCMC run, with the depths from the reflected light plus thermal emission from the upper panel plotted as vertical lines.  The two fitted parameters are eclipse depth (3.83$^{+ 1.10}_{-1.11}$ ppm) and continuum offset from zero (-0.43$\pm$0.41).}
\label{fig:plot1}
\end{figure}
Figure \ref{fig:plot1} shows the result for the list of close-in candidates with $(R_p/a)^2 >$ 10 ppm, containing 5340 individual eclipses, excluding Kepler-10b.  The bin size is approximately 0.0078 in phase, except for the two bins that contain ingress and egress.  Ingress and egress are binned separately, resulting in smaller bin sizes for these two points.  To determine the probability distribution of eclipse depths, we implement a simple Markov Chain Monte Carlo (MCMC) procedure with 500,000 steps, fitting the equation $F_b+ \delta x$, where $F_b$ allows for an offset from zero for the continuum of the light curve, $x$ is the model eclipse curve from \citet{agol}, rescaled so that the continuum is zero and the full eclipse is -1, and $\delta$ is the scale of the eclipse curve.  We use the Bayesian information criterion (BIC, \citealp{bayes}) to establish that the data in this case do not support the addition of a sinusoidal phase curve parameter.  The eclipse depth is then the median value of $\delta$.  Also included in Figure \ref{fig:plot1} is the expected depth of the eclipse calculated from the reflected light plus thermal emission for a range of albedos, adopting $A_{g} = (2/3)A_{Bond}$.  Error bars on the eclipse depth are set by the central 68.27\% of values from the MCMC chain.  The lower panel of Figure \ref{fig:plot1} shows the distributions for the two parameters in the MCMC chain, with the vertical lines indicating the same calculated reflected plus thermal emission values from the upper panel.  We find an average eclipse depth of 3.83$^{+ 1.10}_{-1.11}$ ppm.  The weighted average of $(R_p/a)^2 $ for this group is 17.13 ppm, so if the eclipse is due entirely to reflected light, the average geometric albedo is 0.22 $\pm$ 0.06.  We have excluded Kepler-10b from this group, since it has an eclipse detectable when considered alone \citep{kep10,rouan, new10b,demory}, and the host star is quite bright, at mag$_{kep}$ = 9.12, allowing it to dominate the weighted average.  Note that Kepler-78b does not have short cadence data, so it is not included either.  Including Kepler-10b in the average results in a larger eclipse depth of 5.08$^{+0.71}_{-0.72}$ ppm, consistent with a higher albedo of 0.37 $\pm$ 0.05 if due solely to reflected light.  

Allowing for thermal emission from the planets in the {\it Kepler} bandpass does not change the average geometric albedo appreciably.  Including thermal emission requires adjusting the geometric albedo downward slightly.  We calculated the expected eclipse depth given our average geometric albedo, including thermal emission, which was slightly deeper than the average eclipse depth from the data.  We only need to adjust the geometric albedo downward by 0.001-0.002 to match the average eclipse depth from the data in the case of full redistribution of heat, while we need to adjust it downward by 0.026 to 0.027 to match in the case of instantaneous re-radiation.  These values are less than the uncertainty in the average geometric albedo that we derived from reflected light only.

Our average geometric albedo depends on accurate planet radii.   It is known that cooler stars ($\lesssim$ 4500 K) in the original {\it Kepler} Input Catalog (KIC) have more poorly-determined parameters \citep{kic}, and poorly-determined stellar radii translate to poorly-determined planet radii.  To check how this affects our average geometric albedo, we determined the average geometric albedo for the close-in objects, excluding Kepler-10b, around stars that are $\geq$ 5000 K, and also for those around stars that are $\geq$ 4500 K.   For the 5000 K group, we find an average eclipse depth of 4.25 $^{+ 1.16}_{- 1.15}$ ppm, corresponding to a geometric albedo of 0.25 $\pm$ 0.07.  For the 4500 K group, we find an average eclipse depth of 3.74 $\pm$ 1.11 ppm, corresponding to a geometric albedo of 0.22 $\pm$ 0.06.  Table \ref{tab:depths} summarizes the results for these sub-groupings as well as other sub-groupings which follow.  These are consistent with the result above for the full group, excluding Kepler-10b, so we conclude that the presence of planets orbiting cooler stars in our sample does not present an ambiguity when interpreting our results.

\subsection{Control Group}
\begin{figure}
\epsscale{0.95}
\vspace{0.2 in}
\hspace{0.5 in}
\plotone{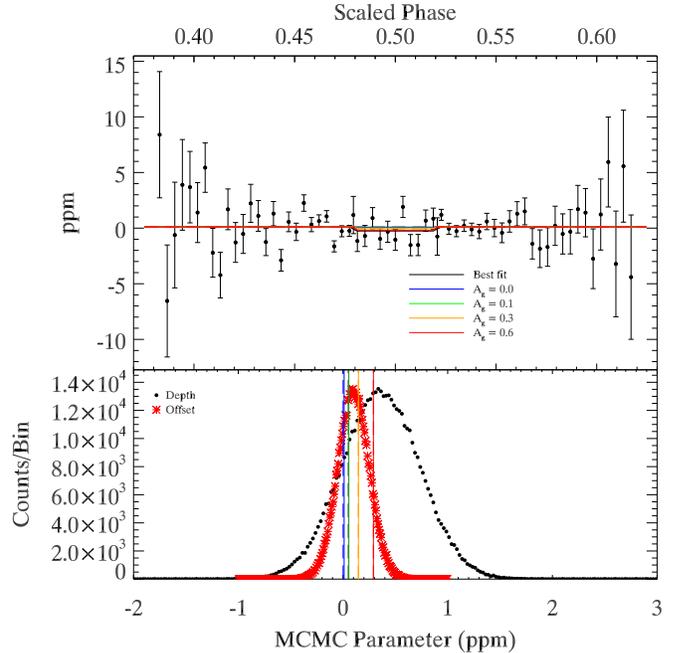}
\vspace{0.4 in}
\caption{Upper panel:  The light curve, centered on secondary eclipse, for the group of objects less than 6 R$_{\earth}$ with $(R_p/a)^2 <$ 1 ppm.  The binned data are shown as points.  The error bars are the propagated photometric errors.  The best fit curve is the solid black line.  Overplotted are the reflected light plus thermal emission models for $A_g = (2/3)*A_B =$ 0.0 (blue), 0.1 (green), 0.3 (orange), and 0.6 (red), with the re-radiation factor $f =$ 1/4 (dashed) and 2/3 (solid).   Lower panel:  The distributions for the two parameters of the MCMC run, with the depths from the reflected light plus thermal emission from the upper panel plotted as vertical lines.  The two fitted parameters are eclipse depth (0.36$\pm$0.37 ppm) and continuum offset from zero (0.10$\pm$0.15 ppm).}
\label{fig:plot3}
\end{figure}

Figure \ref{fig:plot3} shows the averaged light curve, containing 9249 individual eclipses, for the control group of objects with $(R_p/a)^2 <$ 1 ppm.  The bin size is approximately 0.004 in phase.  Using the same two-parameter equation as the close-in group for the MCMC run, we find an average eclipse depth of 0.36 $\pm$ 0.37 ppm, which is consistent with no detection, as expected.  This result cannot constrain the average geometric albedo for this group, which has a weighted average of $(R_p/a)^2 $ = 0.48 ppm. 

To examine the effect that a smaller sample size, similar to that of our close-in group, would have on this result, we chose a subset of the control group randomly.   For 31 candidates, we obtained 3059 eclipses, which resulted in an average eclipse depth of 0.75 $^{+ 0.51}_{- 0.50}$ ppm, again consistent with no detection.  With the full control group, it is impractical to check the stacked eclipses of all of the candidates in the group.  However, for the smaller control group, we found it both practical and important to screen it strongly for false positives, as we did for the close-in group discussed in Sec. 5.1.  This screening identified one system (KOI-116) that we excluded, which we discuss further in Sec. 5.3.

\subsection{Individual Candidates}
\begin{figure}
\epsscale{0.95}
\vspace{0.2 in}
\hspace{0.5 in}
\plotone{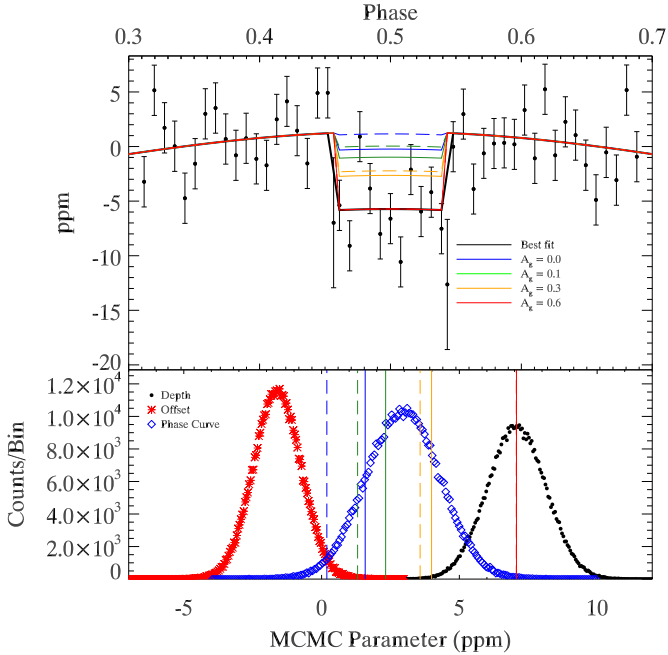}
\vspace{0.4 in}
\caption{Upper panel:  The light curve, centered on secondary eclipse, for Kepler-10b.  The binned data are shown as points.  The error bars are the propagated photometric errors.  The best fit curve is the solid black line.  Overplotted are the reflected light plus thermal emission models for $A_g = (2/3)*A_B =$ 0.0 (blue), 0.1 (green), 0.3 (orange), and 0.6 (red), with the re-radiation factor $f =$ 1/4 (dashed) and 2/3 (solid).   Note that the best fit curve is very similar to the red $A_g$ = 0.6 line and that the ingress and egress bins have large error bars due to the small size of those two bins (cf. Sec. 3).  Lower panel:  The distributions for the three parameters of the MCMC run, with the depths from the reflected light plus thermal emission from the upper panel plotted as vertical lines.  The three fitted parameters are eclipse depth (7.08$\pm$1.06 ppm), continuum offset from zero (-1.61$\pm$0.86 ppm), and semi-amplitude of the phase curve (2.96$\pm$1.36 ppm).}
\label{fig:plot2}
\end{figure}
We consider Kepler-10b on its own, shown in Figure \ref{fig:plot2}, using 898 eclipses from Q2 to Q15, excluding Q8 and Q12.  Data do not exist for these two quarters for this planet because of hardware failure \citep{new10b}.  The bin size is approximately 0.008 in phase.  We do a similar MCMC procedure as for the close-in group, but now we include a third term, the semi-amplitude of a sinusoidal phase curve which peaks at phase 0.5.  This sine curve reproduces the phase curve seen in previous studies of Kepler-10b, and its inclusion is supported by the BIC.  We find a semi-amplitude of 2.96 $\pm$ 1.36 ppm for the phase curve, since we are only fitting from phase 0.25 to phase 0.75, and an eclipse depth of 7.08 $\pm$ 1.06 ppm, giving a geometric albedo of 0.60 $\pm$ 0.09 if due solely to reflected light.  If the semi-amplitude from phase 0.25 to phase 0.75 represents half of the peak-to-peak amplitude of the phase curve, then our eclipse depth and amplitude values are consistent with those found by \citet{kep10,rouan,new10b,demory}, shown in Table \ref{tab:k10bvals}.  Figure \ref{fig:plot2} also shows the distributions for the three parameters in the MCMC chain, as well as the calculated reflected light plus thermal emission eclipse depths for several geometric albedo values.   

\begin{figure}
\epsscale{0.95}
\vspace{0.2 in}
\hspace{0.5 in}
\plotone{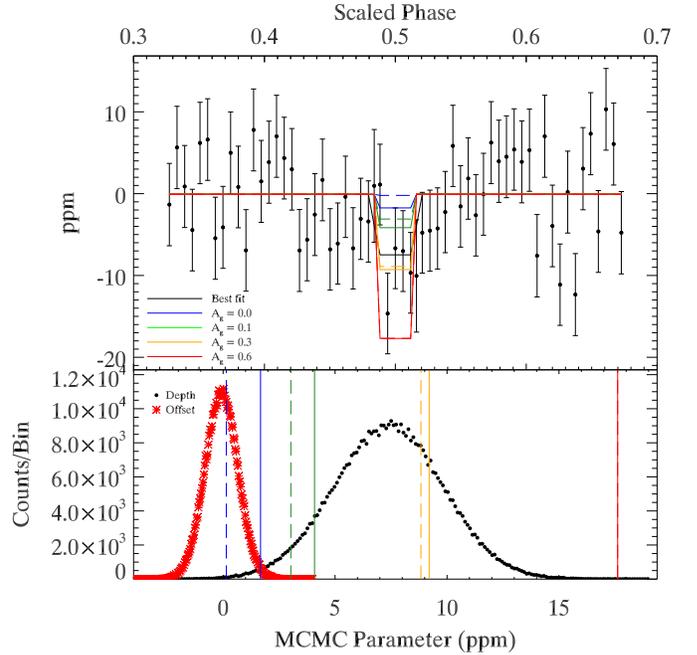}
\vspace{0.4 in}
\caption{Upper panel:  The light curve, centered on secondary eclipse, for KOI-102.01.  The binned data are shown as points.  The error bars are the propagated photometric errors.  The best fit curve is the solid black line.  Overplotted are the reflected light plus thermal emission models for $A_g = (2/3)*A_B =$ 0.0 (blue), 0.1 (green), 0.3 (orange), and 0.6 (red), with the re-radiation factor $f =$ 1/4 (dashed) and 2/3 (solid).   Lower panel:  The distributions for the two parameters of the MCMC run, with the depths from the reflected light plus thermal emission from the upper panel plotted as vertical lines.  The two fitted parameters are eclipse depth (7.40$^{+ 2.42}_{-2.45}$ ppm) and continuum offset from zero (-0.08$\pm$0.73 ppm).}
\label{fig:koi102}
\end{figure}
We find that KOI-102.01 also has a significant, physically plausible detection at 7.40$^{+ 2.42}_{-2.45}$ ppm, shown in Figure \ref{fig:koi102}.   We keep this candidate in the average for Figure \ref{fig:plot1} because at mag$_{kep}$ = 12.57, it does not dominate the weighted average.  {For the individual analysis of this candidate, we include data that is $\pm$ 5 times the duration of the full eclipse, centered on phase 0.5 to improve the baseline.  We also change the bin size to approximately 0.0058 in phase, such that there are 5 bins within full eclipse.  The candidate has $(R_p/a)^2$ = 29.36 ppm, resulting in a geometric albedo of 0.25 $\pm$ 0.08 if due only to reflected light.  With $R_p$ = 3.69 $R_{\earth}$, this albedo puts the candidate in a class of reflective hot Neptunes.

KOI-676.02 may have an eccentric orbit.  \citet{kep210} suggest the most probable configuration for the KOI-676 (Kepler-210) system is that the inner planets, KOI-676.01 and KOI-676.02, have a large eccentricity, based on the mismatch of $a/R_*$ between the two.  A third member of the system is hypothesized from transit timing variations (TTVs).  The suggested configuration has the major axis of the orbit of KOI-676.02 along the line-of-sight to Earth.  We include KOI-676.02 in our close-in group at the moment but will remove it if the eccentricity is confirmed.  Note that KOI-676 has $T_{eff} <$ 4500 K, so this candidate was not included in the averages above for stars with $T_{eff} >$ 5000 K and $T_{eff} >$ 4500 K.

\begin{figure}
\epsscale{0.95}
\vspace{0.2 in}
\hspace{0.5 in}
\plotone{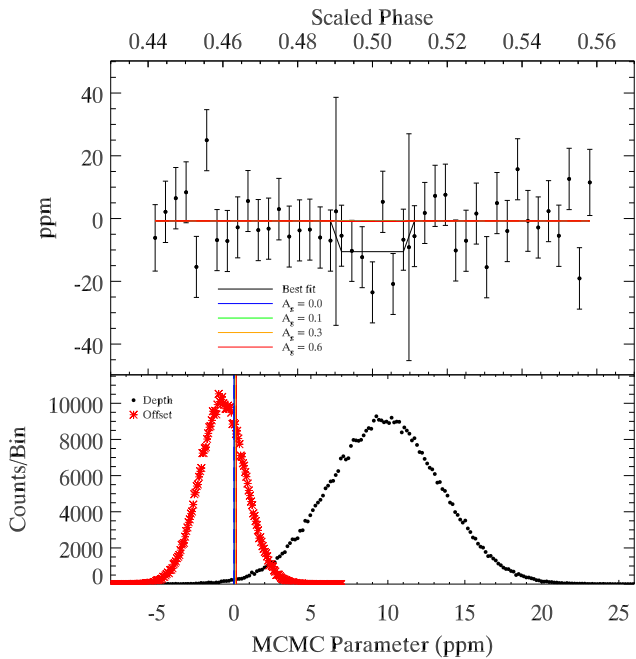}
\vspace{0.4 in}
\caption{Upper panel:  The light curve, centered on secondary eclipse, for KOI-116.03.  The binned data are shown as points.  The error bars are the propagated photometric errors.  The best fit curve is the solid black line.  Overplotted are the reflected light plus thermal emission models for $A_g = (2/3)*A_B =$ 0.0 (blue), 0.1 (green), 0.3 (orange), and 0.6 (red), with the re-radiation factor $f =$ 1/4 (dashed) and 2/3 (solid).  Note that the ingress and egress bins have large error bars due to the small size of those two bins (cf. Sec. 3).  Lower panel:  The distributions for the two parameters of the MCMC run, with the depths from the reflected light plus thermal emission from the upper panel plotted as vertical lines.  The two fitted parameters are eclipse depth (9.80$^{+ 3.64}_{- 3.65}$ ppm) and continuum offset from zero (-0.75$\pm$1.48 ppm).}
\label{fig:116}
\end{figure}
We exclude KOI-116 (Kepler-106) from our control group.  The original control group contained KOI-116.01 (Kepler-106c), KOI-116.03 (Kepler-106b), and KOI-116.04 (Kepler-106d).  When considering the shortened control group, we found that KOI-116.03 has a 2-$\sigma$ detection of an eclipse on its own, 9.80$^{+ 3.64}_{- 3.65}$ ppm, which is far deeper than expected even for a high geometric albedo.   We use a bin size of approximately 0.0028 in phase, which corresponds to 7 bins inside of full eclipse.  The candidate has $(R_p/a)^2$ = 0.273 ppm.  If the eclipse were solely due to reflected light, this would require a non-physical geometric albedo greater than 30.  The candidate has a maximum equilibrium temperature of 1434 K, assuming $A_B$ = 0 and instantaneous re-radiation, so thermal emission due to stellar irradiation cannot explain the large eclipse depth either.  Although the evidence for the planetary nature of KOI-116.03 seems solid \citep{marcy},  our stacked eclipse data suggest that there may be other effects at play in this system.  The nature of those effects is beyond the scope of this paper, but we show our stacked eclipses for KOI-116.03 in Figure \ref{fig:116}.

\subsection{Advantages and Limitations of the Technique}

Our average is biased towards brighter, and hence closer, stars, due to the weighting by photometric noise.  There is no reason to expect a correlation between the stellar distance from Earth and planet type or albedo, so we do not expect this bias to favor bright or dark planets.  The case of Kepler-10b shows that we must be careful in our sample selection, however, when the sample size is small, because an atypically bright planet around an atypically bright star in the sample can alter the results.  Likewise, false positives could have a bigger influence when the sample size is small, so it is important to screen the individual candidates for any obvious signs that they could be strongly influencing the result.  To limit the risk of including false positives, we only considered objects deemed to be ``planet candidates'', which means that the candidate has not failed any of the false positive tests in \citet{bat13} applied thus far.  We also examined the stacked eclipses of each candidate individually, to discriminate against non-physically large effects such as we find for KOI-116.03.

The major disadvantage of our technique is that the we cannot tell whether the average represents a typical object in the group.  The underlying distribution could be bimodal, with the average representing a non-existent object in between the two groups.  Another disadvantage is that our eclipse depth could be diluted due to astrophysical effects that we cannot account for at this time.  If any of the planet candidates are in eccentric orbits, their eclipses would not occur at the expected phase and would destructively add to the signal.  False positives could also dilute the signal, because there may be no eclipse to detect.  For example, eclipsing binaries in which the two stars are of similar type could have primary and secondary eclipses that are approximately equal.   If the binary is blended with the KOI star, both the primary and the secondary eclipses would look like planetary transits at half the period of the binary, and, if included in the average, the object would add in a flat signal when looking at phase 0.5 based on the planetary interpretation \citep[e.g.][]{borucki}.  The referee also points out that cases in which we see a planetary transit but do not see the eclipse due to the geometry of system would result in extra light at the expected eclipse, creating a bump in the light curve that would decrease our average signal.  This scenario would only occur if the orbits are not circular, and we expect that the orbits of the close-in candidates have circularized.  Uncertainties in the host star parameters could also be detrimental, as the planet radius is tied directly to the stellar radius when measured using the transiting technique.   Changing the size of a planet would affect our calculated albedo, since the calculation depends on the planet radius.  

In spite of its limitations, an overwhelming advantage of our technique is that we reach a much greater signal-to-noise ratio in the eclipse light curve than is achievable for many of these planet candidates individually.   Moreover, we can use a large sample of planet candidates to obtain information about the average albedo without being biased towards the candidates that are most easily measured, which tend to be more reflective than average.  Future space missions such as TESS \citep{ricker} and PLATO \citep{rauer} will discover even greater numbers of transiting planets, many of which will be too faint to allow individual characterization of their atmospheres.  Grouping planets to determine their average atmospheric properties will become an increasingly relevant and important tool in the future. 

\section{Implications for Atmospheres and Surfaces of Sub-Saturns}

\subsection{Presence of Atmospheres in Our Sample}
\citet{heng} model the stability of atmospheres on tidally-locked Earth-like planets, where the body is mostly rocky with a thin atmosphere of varying mean molecular weights ($\mu$).  They consider the stability against condensation on the night-side of the planet, for planets around G, K, and M dwarfs.  They place {\it Kepler} candidates up to 6 $R_{\earth}$ on these stability diagrams in their Figures 3 and 6.  Two of our close-in candidates have F dwarf hosts, but we can locate the remaining 30 candidates around G, K, and M dwarfs on their stability diagrams.  All but one candidate lie in the stable region if the atmospheres have low $\mu$, but none are stable with an Earth-like atmospheric composition.  The single candidate outside the stable region even at low $\mu$ is KOI-356.01, though it is very close to the boundary.  

Many of our close-in candidates, however, are likely to be more Neptune-like than Earth-like.  \citet{lopez13} suggest a cut-off radius for rocky planets of 1.75 $R_{\earth}$, based on the radius-composition relation they find in their planet formation models.  \citet{rogers} sets the bar even lower, determining a cut-off of 1.6 $R_{\earth}$ through a hierarchical Bayesian analysis of {\it Kepler} planets with mass limits determined from radial velocity studies.  Of our 32 close-in candidates, 12 would be considered rocky based on these cut-off values.  KOI-356.01 would be more Saturn-like, at a radius of 5.73 $R_{\earth}$.   There remains a small probability that the planet candidates above 1.6 $R_{\earth}$ could be rocky.  Kepler-10c, despite being 2.35 $R_{\earth}$, has a density of 7.1 g cm$^{-3}$ \citep{k10c}, suggesting a rocky nature with a significant amount (5 to 20 wt.\%) of water or some other high $\mu$ volatile.  The candidates smaller than 1.5 $R_{\earth}$ are quite likely to be rocky, while the candidates above 2.5 $R_{\earth}$ are likely Neptune-like with substantial atmospheres, and KOI-299.01, at 1.98 $R_{\earth}$, may fall somewhere in between the two types, like Kepler-10c.

\subsection{Expected Albedos}
\begin{figure}
\epsscale{1.25}
\hspace{-0.3 in}
\plotone{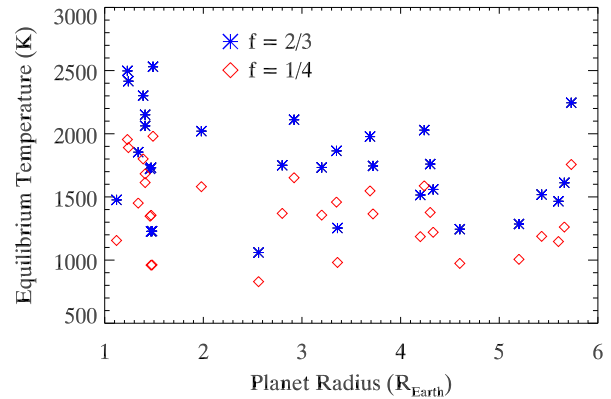}
\vspace{-0.2 in}
\caption{The calculated equilibrium temperatures of the close-in group, excluding Kepler-10b, plotted versus planet radius.  The equilibrium temperatures are calculated assuming our average geometric albedo and Lambert's law.  The blue symbols use $f=2/3$, for instantaneous re-radiation, while the red symbols use $f=1/4$, for complete redistribution of heat.}
\label{fig:teqvsrp}
\end{figure}
The candidates in our close-in group in Sec. 5.1 are too hot to have water clouds in their upper atmospheres \citep{morley}.  Figure \ref{fig:teqvsrp} shows the equilibrium temperature, calculated using our average geometric albedo, versus planet radius for all of the objects in the close-in group, excluding Kepler-10b.  If the candidates all have poor heat redistribution (i.e. $f = 2/3$), they are all above 1000 K.   Thus the albedo we determine cannot be due to water clouds.

For giant planets, \citet{sudarsky} show that alkali metals like Na strongly absorb visible light in atmospheres that are too hot for water clouds but cool enough that a silicate layer does not form.   Above 900 K, a silicate layer forms, but it forms at low enough altitudes that the alkali metals above the layer still absorb a significant amount of the incident visible light, producing the lowest albedos.  As the temperature increases, the silicate layer forms higher up in the atmosphere, potentially reflecting the incident light before it can interact with the alkali metals, producing much higher albedos.  The Bond albedo due to silicates could be as high as 0.55, which translates to a geometric albedo of 0.37 for Lambertian surfaces.   \citet{demoryetal} find $A_g = 0.32 \pm 0.03$ for Kepler-7b, which agrees with the predictions of \citet{sudarsky} for reflective cloud layers, but they find that the high albedo can also be explained solely with Rayleigh scattering if the planet's atmosphere is depleted in alkali metals relative to solar abundances by a factor of 10-100.   Other hot Jupiters with similiar equilibrium temperatures, such as HD 209458b, have been found to be very dark, with upper limits on the geometric albedo of 0.12 \citep{rowe08}, suggesting that if reflective clouds produce the high albedo of Kepler-7b, then there must be some mechanism that suppresses the cloud formation in HD 209458b and other dark hot Jupiters.

\citet{miguel} model rocky planets at these temperatures with tenuous atmospheres due to vaporization of the surface.   For atmospheres up to nearly 2900 K, monatomic Na dominates the atmosphere, with the fraction of SiO increasing with temperature, and, by 2900 K, the atmosphere becomes dominated by SiO and monatomic Na.   If the smaller planets can also form a silicate layer high in their atmospheres, that could boost their albedos.   Alternatively, if their atmospheres are tenuous enough, the light may be reflecting off the surfaces such as in the ``lava ocean'' model proposed for Kepler-10b and Kepler-78b \citep{leger2011,rouan,sanchis}. 

Our average albedo for the close-in sub-Saturns suggests that the high albedos of Kepler-10b and Kepler-78b are atypical and that sub-Saturns are typically more reflective than hot Jupiters.  Limiting our close-in group to planets $< 2 R_{\earth}$, excluding Kepler-10b, results in a non-detection of an eclipse with a depth of 2.14 $\pm$ 1.96 ppm, using an MCMC trial similar to those described above.  This result gives less than a 1\% chance of an average geometric albedo $\ge$ 0.6 and a 21\% chance of an average geometric albedo $\ge$ 0.3.  Moreover, thermal emission alone cannot explain the eclipse depth of Kepler-10b, unless the planet has some source of internal heat or a spectrum that is not a blackbody curve \citep{new10b}.  If the ``lava ocean'' model proposed for Kepler-10b and Kepler-78b is indeed the source of the high albedo needed to match the eclipse depths for these two planets, then close-in super-Earths more typically have darker surfaces, or absorption due to alkali metals or hazes in their tenuous atmospheres.  They could also have mildly reflective clouds that shield the more reflective lava ocean surface.   

Limiting our close-in group to the planets $> 2 R_{\earth}$ results in an average eclipse depth of 4.68 $\pm$ 1.22 ppm, giving a geometric albedo of 0.23 $\pm$ 0.06.  The Neptune-like sub-Saturns likely also have clouds or hazes that reflect the incident light back before being substantially absorbed by the alkali metals in their atmospheres.  Alternatively, the average eclipse depth we find could be due to thermal inversions in the Neptune-like atmospheres, as have been seen in and modeled for some hot Jupiters \citep[e.g.][]{fortney,zahnle}.  \citet{spiegel} note that Neptune and Uranus have much higher metallicities than Jupiter, and so Neptune-like planets could have extra absorbers due to the increased metals that could aid the creation of thermal inversions.  If a thermal inversion exists, the thermal emission contribution to the eclipse depths could be greater, reducing the albedo necessary to match the depths. 

\section{Summary}

We average eclipses of 31 close-in planet candidates $< 6 R_{\earth}$ to determine the average albedo of the group.  We find that, on average, close-in sub-Saturns are not extremely dark, with an average geometric albedo of 0.22.  This albedo is consistent with the results for close-in super-Earths by \citet{demory}, and it is in contrast to many hot Jupiters, which have albedos $<$ 0.1, and to Kepler-10b and Kepler-78b, which have albedos of 0.4-0.6.  The super-Earths may have darker surfaces than Kepler-10b and Kepler-78b, or they may have clouds or hazes in their tenuous alkali metal atmospheres that either absorb some of the light or reflect the light before it can reach the higher albedo surface.  The Neptune-like planets may also have reflective clouds or hazes, preventing absorption lower in their atmospheres by the alkali metals that produce the very low albedos in hot Jupiters.   The Neptune-like planets may also have thermal inversions that add extra thermal emission to the eclipse depths, making the albedo appear higher than it is.

\section{Acknowledgements}
We would like to thank the referee for comments and suggestions which helped to improve this work.  This paper includes data collected by the Kepler mission. Funding for the Kepler mission is provided by the NASA Science Mission directorate.  All of the data presented in this paper were obtained from the Mikulski Archive for Space Telescopes (MAST). STScI is operated by the Association of Universities for Research in Astronomy, Inc., under NASA contract NAS5-26555. Support for MAST for non-HST data is provided by the NASA Office of Space Science via grant NNX13AC07G and by other grants and contracts.  This research has made use of the NASA Exoplanet Archive, which is operated by the California Institute of Technology, under contract with the National Aeronautics and Space Administration under the Exoplanet Exploration Program.

\begin{deluxetable}{lccccccc}
\tabletypesize{\footnotesize}
\tablecolumns{8}
\tablewidth{0pt}
\tablecaption{Candidate Parameters \label{tab:params}}
\tablehead{
\colhead{KOI} & \colhead{\# eclipses} & \colhead{$R_p$\tablenotemark{a}} & \colhead{$\left(R_p/a\right)^2$\tablenotemark{a}} & \colhead{$a/R_*$\tablenotemark{a}} & \colhead{Max $T_{eq}$\tablenotemark{b}} & \colhead{Min $T_{eq}$\tablenotemark{c}} & \colhead{$T_{eff}$} \\
& & \colhead{$\left(R_{\earth}\right)$} & \colhead{(ppm)} & & \colhead{(K)} & \colhead{(K)} & \colhead{(K)}
}
\startdata
\hline
\multicolumn{8}{l}{Kepler-10b} \\
\hline
K00072.01 & 898 &  1.37 &  11.78 &    3.7 & 2659.0 & 1170.1 &   5627 \\
\hline
\multicolumn{8}{l}{$\left(R_p/a\right)^2 > 10$ ppm} \\
\hline
K00005.01 & 233 &  5.66 &  17.27 &    8.8 & 1786.8 &  786.3 &   5861 \\
K00007.01 & 249 &  3.72 &  12.96 &    7.5 & 1939.0 &  853.3 &   5858 \\
K00046.01 & 79 &  4.33 &  15.39 &    9.1 & 1721.9 &  757.8 &   5764 \\
K00102.01\tablenotemark{d} & 556 &  3.69 &  29.36 &    5.8 & 2194.0 &  965.5 &   5838 \\
K00104.01 & 309 &  3.36 &  17.71 &    9.7 & 1391.4 &  612.3 &   4786 \\
K00141.01 & 256 &  5.43 &  39.06 &    8.5 & 1678.4 &  738.6 &   5425 \\
K00191.03 & 547 &  1.24 &  12.39 &    3.7 & 2688.0 & 1182.9 &   5696 \\
K00240.01 & 18 &  4.20 &  11.39 &   11.2 & 1679.8 &  739.2 &   6215 \\
K00299.01 & 473 &  1.98 &  10.52 &    5.0 & 2229.4 &  981.1 &   5538 \\
K00356.01 & 11 &  5.73 &  75.96 &    3.8 & 2498.2 & 1099.4 &   5364 \\
K00433.01 & 55 &  4.60 &  18.14 &   11.9 & 1377.1 &  606.0 &   5262 \\
K00505.03 & 31 &  3.35 &  11.54 &    4.9 & 2073.8 &  912.6 &   5058 \\
K00676.02 & 290 &  2.56 &  13.21 &   11.3 & 1171.8 &  515.6 &   4367 \\
K00697.01 & 123 &  4.24 &  17.63 &    5.4 & 2238.8 &  985.2 &   5779 \\
K00739.01 & 121 &  1.48 &  11.00 &    7.6 & 1361.6 &  599.2 &   4153 \\
K00755.01 & 13 &  2.80 &  10.97 &    7.7 & 1933.1 &  850.7 &   5953 \\
K00800.01 & 21 &  3.20 &  12.21 &    8.4 & 1920.9 &  845.3 &   6157 \\
K00936.02 & 71 &  1.47 &  17.42 &    6.5 & 1358.3 &  597.7 &   3834 \\
K01128.01 & 37 &  1.41 &   9.99 &    4.7 & 2295.2 & 1010.0 &   5480 \\
K01169.01\tablenotemark{e} & 41 &  1.49 &  15.73 &    3.4 & 2805.1 & 1234.4 &   5719 \\
K01239.01 & 53 &  1.39 &  12.13 &    4.3 & 2548.2 & 1121.3 &   5849 \\
K01300.01 & 336 &  1.34 &  22.62 &    4.1 & 2041.3 &  898.3 &   4602 \\
K01367.01 & 236 &  1.41 &  25.04 &    3.7 & 2384.1 & 1049.1 &   5070 \\
K01428.01 & 253 &  1.46 &  13.38 &    5.3 & 1899.9 &  836.1 &   4858 \\
K01442.01 & 690 &  1.23 &  12.20 &    3.2 & 2749.2 & 1209.8 &   5476 \\
K01510.01 & 40 &  1.47 &  15.31 &    5.3 & 1924.2 &  846.8 &   4885 \\
K01784.01 & 13 &  5.20 &  15.09 &   13.8 & 1424.9 &  627.1 &   5853 \\
K01805.01 & 51 &  5.60 &  11.61 &   10.1 & 1622.5 &  714.0 &   5708 \\
K01835.02 & 51 &  2.92 &  14.20 &    3.9 & 2325.2 & 1023.2 &   5110 \\
K02678.01 & 8 &  4.30 &  15.85 &    6.2 & 1945.5 &  856.1 &   5371 \\
K02700.01 & 75 &  1.12 &   8.89 &    6.0 & 1635.9 &  719.9 &   4433 
\enddata
\tablenotetext{a}{$R_p$, $a$, and $R_*$ are taken from the Exoplanet Archive candidate table, and $(R_p/a)^2$ and $a/R_*$ are calculated from them.}
\tablenotetext{b}{Assumes $f = 2/3$ (instant re-radiation) and $A_B = 0.0$.}
\tablenotetext{c}{Assumes $f=1/4$ (complete redistribution) and $A_B = 0.9$ (i.e. $A_g = 0.6$).}
\tablenotetext{d}{\protect{\citet{coughlin}} find an eclipse depth of 13.6$^{+8.83}_{-9.86}$ ppm, consistent with the depth of 7.43$^{+2.32}_{-2.30}$ found in this work.}
\tablenotetext{e}{\protect{\citet{demory}} finds an eclipse depth of 13.5$^{+3.0}_{-3.1}$ ppm, using long cadence data.  With the more limited amount of short cadence data, we find a depth of 20.31$^{+14.31}_{-14.35}$ ppm, consistent with the higher-precision measurement.}
\end{deluxetable}

\begin{deluxetable}{lcc}
\tabletypesize{\footnotesize}
\tablecolumns{3}
\tablewidth{0pt}
\tablecaption{Eclipse depths and albedos \label{tab:depths}}
\tablehead{\colhead{Group} & \colhead{Avg Eclipse Depth} & \colhead{Avg $A_g$} \\
& \colhead{(ppm)} & 
}
\startdata
\hline
\multicolumn{3}{l}{Close-in, $(R_p/a)^2 > 10$ ppm} \\
\hline
without Kepler-10b & 3.83 $^{+1.10}_{-1.11}$ & 0.22$\pm$0.06 \\
with Kepler-10b & 5.08 $^{+0.71}_{-0.72}$ & 0.37$\pm$0.05 \\
$T_{eff} > 5000$ K & 4.25$^{+1.16}_{-1.15}$ & 0.25$\pm$0.07\\
$T_{eff} > 4500$ K & 3.74$\pm$1.11 & 0.22$\pm$0.06\\
$R_p < 2 R_{\earth}$ \tablenotemark{a} & 2.14$\pm$1.96 & 0.17$\pm$0.16\\
$2 R_{\earth} < R_p < 6 R_{\earth}$ & 4.68$\pm$1.22 & 0.23$\pm$0.06 \\
\hline
\multicolumn{3}{l}{Control, $(R_p/a)^2 < 1$ ppm}\\
\hline
376 candidates & 0.36$\pm$0.37 & 0.75$\pm$0.77 \\
31 candidates & 0.75$^{+0.51}_{-0.50}$ & 1.24$^{+0.85}_{-0.83}$\\
\hline
\multicolumn{3}{l}{Individual Candidates}\\
\hline
Kepler-10b & 7.08$\pm$1.06 & 0.60$\pm$0.09 \\
KOI-102.01 & 7.43$^{+2.32}_{-2.30}$ & 0.25$\pm$0.08\\
KOI-116.03 & 8.04$^{+3.71}_{-3.67}$ & \ldots \tablenotemark{b}
\enddata
\tablenotetext{a}{Excludes Kepler-10b.}
\tablenotetext{b}{The calculated geometric albedo is unphysical.}
\end{deluxetable}

\begin{deluxetable}{lcc}
\tabletypesize{\footnotesize}
\tablecolumns{3}
\tablewidth{0pt}
\tablecaption{Kepler-10b Studies \label{tab:k10bvals} }
\tablehead{
\colhead{Authors} & \colhead{Phase Curve Amplitude} & \colhead{Eclipse Depth} \\
& \colhead{(ppm)} & \colhead{(ppm)}
}
\startdata
Batalha et al. (2011) & 7.6$\pm$2.0 & 5.8$\pm$2.5 \\
Rouan et al. (2011) & 5.6$\pm$2.0 & 5.6$\pm$2.0 \\
Fogtmann-Schulz et al. (2014) & 8.13$\pm$0.68 & 9.91$\pm$1.01 \\
Demory (2014) & $\cdots$ & 7.4$^{+1.1}_{-1.0}$ \\
This Work & 5.92$\pm$2.72 & 7.08$\pm$1.06 
\enddata
\end{deluxetable}

\end{document}